\DocumentMetadata{}
\documentclass[sigconf,authorversion,natbib=true,nonacm]{acmart}

\AtBeginDocument{%
  }

\usepackage{multirow}
\usepackage{amsmath}

\usepackage{relsize}
\usepackage{graphicx}
\usepackage{listings}
\usepackage{xcolor}
\usepackage{booktabs}
\usepackage{fancyref}
\usepackage{todonotes}
\usepackage{import}
\usepackage{dsfont}
\usepackage{lineno}
\usepackage{float}
\usepackage{caption}
\usepackage[skip=10pt]{subcaption}
\usepackage{bm}
\usepackage{bbm}
\usepackage{commath}
\usepackage[section]{placeins}
\usepackage{fancyhdr}
\usepackage{natbib}
\usepackage{datetime}
\usepackage{booktabs}

\usepackage{algorithmic}
\usepackage[ruled,vlined]{algorithm2e}
\usepackage{makecell}
\usepackage{ifthen}
\usepackage{makecell}
\usepackage{siunitx}
\usepackage{mathtools} 
\mathtoolsset{showonlyrefs=true} 
\usepackage{amsthm}
\usepackage{bbm}

\usepackage{varwidth}
\usepackage{capt-of}

\usepackage{tikz}
\usetikzlibrary{calc,matrix,chains,positioning,decorations.pathreplacing,arrows}

\newdate{date}{1}{1}{2023}

\newcommand{\one}{\mathbbm{1}}
\newcommand{\N}{\mathbb{N}}
\newcommand{\R}{\mathbb{R}}

\newcommand{\cT}{\mathcal{T}}

\newcommand{\<}{\langle}
\renewcommand{\>}{\rangle}

\theoremstyle{definition} 
\begin{document}

\title[NeuralBeta: Estimating Beta Using Deep Learning]{NeuralBeta:\\Estimating Beta using Deep Learning}

\author{Yuxin Liu}
\affiliation{%
  \institution{Bloomberg}
  \streetaddress{731 Lexington Ave}
  \city{New York}
  \country{USA}}
\email{yliu1754@bloomberg.net}

\author{Jimin Lin}
\affiliation{%
  \institution{Bloomberg}
  \streetaddress{731 Lexington Ave}
  \city{New York}
  \country{USA}}
\email{jlin846@bloomberg.net}

\author{Achintya Gopal}
\affiliation{%
  \institution{Bloomberg}
  \streetaddress{731 Lexington Ave}
  \city{New York}
  \country{USA}}
\email{agopal6@bloomberg.net}



\begin{abstract}
Traditional approaches to estimating $\beta$ in finance often involve rigid assumptions and fail to adequately capture $\beta$ dynamics, limiting their effectiveness in use cases like hedging. To address these limitations, we have developed a novel method using neural networks called \textit{NeuralBeta}, which is capable of handling both univariate and multivariate scenarios and tracking the dynamic behavior of $\beta$. To address the issue of interpretability, we introduce a new output layer inspired by regularized weighted linear regression, which provides transparency into the model's decision-making process. We conducted extensive experiments on both synthetic and market data, demonstrating \textit{NeuralBeta}'s superior performance compared to benchmark methods across various scenarios, especially instances where $\beta$ is highly time-varying, e.g., during regime shifts in the market. This model not only represents an advancement in the field of beta estimation, but also shows potential for applications in other financial contexts that assume linear relationships.
\end{abstract}

\maketitle

\section{Introduction}

The term “beta”, or $\beta$, represents the linear coefficient between an explanatory variable and a response variable, and plays a crucial role in finance markets for asset pricing, portfolio optimization, and risk management. In asset pricing, $\beta$ is used to evaluate the return of an asset based on, for example, the return of the market portfolio (Capital Asset Pricing Model \citep{sharpe1964capital}), market plus asset characteristics (Fama-French three-factor model and five-factor model \cite{fama1993common, fama2015five}), or a set of risk factors and market indices (Arbitrage Pricing Theory \citep{ross2013arbitrage}).

The original CAPM model assumes that stock market betas remain constant over time. However, empirical research including \cite{bollerslev1988capital} and \cite{jagannathan1996conditional} demonstrates that betas are indeed time-varying. Traditional methods to model this time variation usually involve rolling-window regressions using Ordinary Least Squares (OLS) and Weighted Least Squares (WLS). But these approaches face challenges in specifying the window length and weights to balance bias and variance. These regression methods are also sensitive to outliers, which leads to volatile beta estimates.

In recent years, with the proliferation of machine learning in finance, research applying machine learning techniques to beta estimation remains sparse. \cite{drobetz2023estimating}, for example, utilizes future information to compute the “true beta” (or realized beta) as the target and evaluates several machine-learning based estimators against this metric, and they found random forests to be the most effective. A limitation of this approach is the introduction of new hyperparameters in computing realized beta (the window length, the weights, etc.) and that using future data in computing the target reduces the amount of data available. Moreover, \cite{alanis2024can} explores the use of machine learning algorithms to estimate equity betas from historical returns and company financials, for which they also used the realized beta as the target. They focus primarily on private firms or non-traded assets, which diverges from our area of interest. Notably, none of these approaches employ state-of-the-art deep neural networks like attention, nor do they provide the transparency needed in financial modelling. The usage of realized beta as the target is also questionable, as the primary usage of beta is hedging, and targeting a beta using a future window does not necessarily improve this task.

Our paper contributes to the literature by proposing an interpretable approach to beta estimation using deep neural networks, which we call \textit{NeuralBeta}. \textit{NeuralBeta} directly uses the hedging error as the metric, and leverages the power of neural networks to capture complex, non-linear relationships in financial data, offering more precise and robust estimation of $\beta$. We also develop an interpretable variant, \textit{NeuralBeta-Interpretable} (NBI). NBI enhances transparency by explicitly outputting weights for each data point in the lookback window, which are then used to calculate the weighted least squares (WLS) solution to estimate $\beta$. This approach allows users to easily discern which historical periods are more influential in the beta estimation process and to identify specific temporal patterns and market conditions that influence beta. 

We highlight the following threefold advantages of \textit{NeuralBeta}:

\textbf{Generality}: \textit{NeuralBeta} utilizes neural networks to cover a large set of possible functional forms of beta. It overcomes the limitation of conventional approaches that train models separately on small sub-datasets, by training jointly on the entire dataset of all assets and factors. Hence it can better capture both general and specific patterns.

\textbf{Interpretability}: A novel interpretable architecture is constructed to explicitly quantify the data importance in the form of sample weighting, which is directly related to the traditional WLS estimation framework. This feature allows users to analyze which data points are considered most critical by the model and to gain insights into how the model works.

\textbf{Practicality}: \textit{NeuralBeta} employs Mean Squared Error (MSE) on asset returns as the evaluation metric, which is a hedging performance-based measure. Compared to metrics based on realized beta versus predicted beta, the hedging performance metric is more relevant to investment performance in practice.

The structure of this paper is as follows. In Section \ref{sec:methodology}, we introduce the general framework of beta estimation and the \textit{NeuralBeta} model architecture, both the non-interpretable and interpretable versions. In Section \ref{sec:experiment}, we apply \textit{NeuralBeta} on both synthetic and market data and compare it with benchmark methods. 
We show that \textit{NeuralBeta} systematically outperforms other conventional methods, and that our interpretable NBI architecture achieves the balance between explainability and performance.
We also provide examples of the weights outputted by NBI. Finally, we summarize our findings in Section \ref{sec:conclusion}.

\section{Methodology}\label{sec:methodology}

\subsection{Problem Setup}

To develop the methodology, we start from a simple scenario where one hedges a single target asset against multiple other hedging instruments on a daily basis. The task is to determine the optimal hedging ratio of each instrument that minimizes the next day hedging error, and ultimately to achieve a low average hedging error over the entire horizon. In the following we will formalize the hedging task. With the single target hedging scenario formulated, we will see it easily scales up to multiple assets case and applies to any other prediction tasks.

Let $\cT = \{0, 1, 2, \dots\}$ be the discrete time index for data of certain frequency. We refer to the frequency in terms of days. For a time interval $(s, t]$, $s, t \in \cT$ and $s < t$, let $D_{s,t} := \{(x_{s+1}, y_{s+1}), \allowbreak (x_{s+2}, y_{s+2}), \dots, (x_t, y_t)\}$ denote the dataset sliced between time $s$ and time $t$, where $x \in \R^d$ is the explanatory variable of dimension $d \in \N$, e.g., factor returns, and $y \in \R$ is the scalar response variable, e.g., a single stock return.

For any $t \in \cT$, we are interested in approximating a linear relationship between $x_t$ and $y_t$ with coefficient $\beta_t \in \R^d$ and noise $\epsilon_t \in \R$, which is of the form
\begin{equation} \label{eq:beta_estimation}
y_t= \<\beta_t, x_t\> + \epsilon_t 
\end{equation}
where $\<\cdot, \cdot\>$ is the inner product.

A one-step hedging task at time $t$ is to determine an optimal hedging ratio $\hat{\beta}_{t+1}$ given available data $D_{0, t}$ such that the ex-ante hedging error at time $t+1$ is minimized:
\begin{align} \label{eq:obj_t}
\hat{\beta}_{t+1} = \min_\beta L\left(y_{t+1}, \<\beta, x_{t+1}\>\right),
\end{align}
where $L$ is some risk measure, such as expected quadratic loss or negative log-likelihood. Here, $\hat{\beta}_{t+1}$ is assumed to be inferable from the dataset $D_{0,t}$, i.e., it is in the form
\begin{align} \label{eq:beta_f}
\hat{\beta}_{t+1} = f(t, D_{0,t})
\end{align}
for some function $f:\cT \times D \to \R^d$ that can possibly be time inhomogeneous, which is an estimator given by the underlying model.

Then, we consider the following practical scenario. During a $n$-day time horizon starting at day $\tau \in \cT$ denoted by $\cT_\tau^n = \{\tau, \tau+1 , \dots \tau + n - 1\}$ with $\tau \ge h$ for some chosen lookback window $h \in \N$ (such that there are at least $h$ historical data points at $\tau$). We perform the one-step hedge ratio prediction with \eqref{eq:beta_f} at each day $t \in \cT_\tau^n$.
The objective is thus naturally to find the function $f$ that minimizes the average hedging error through the entire time horizon:
\begin{align} \label{eq:obj}
\min_f \frac{1}{n} \sum_{t \in \cT_\tau^n} L\left(y_{t+1}, \< \hat{\beta}_{t+1}, x_{t+1}\>\right)
\end{align}

It is important to distinguish objective \eqref{eq:obj} from the conventional regression task. First, the entire dataset available to each one-step objective is streamed progressively as $D_{0, \tau}, D_{0, \tau+1}, \dots$ instead of a static full dataset. Second, in each day $t \in \cT_\tau^n$, we are interested in estimating the coefficient that minimizes the next day error revealed by the new data point $(x_{t+1}, y_{t+1})$, instead of the coefficient that minimizes the hedging error in the past. Third, we don't presume the ground truth coefficient $\beta$ to be time-homogeneous, so the coefficient is expressed in a functional form as in equation \eqref{eq:beta_f} instead of being sets of constants.

Nevertheless, we can still draw links from objective \eqref{eq:obj} as a meta-task to a static regression task, or a rolling regression task in the following sense.

\paragraph{Ordinary Linear Regression} Suppose we do a one-step hedge. At $\tau$, if the data $D_{0, \tau+1}$ satisfies the Gauss-Markov assumptions, then the optimal functional form of $\beta$ is given by
\begin{align} \label{eq:ols}
f^*(\tau, D_{0, \tau}) = f^*(D_{0, \tau}) = \left(X_{0, \tau}^T X_{0, \tau}\right)^{-1}X_{0, \tau}^T y_{0, \tau},
\end{align}
where $X_{0, \tau}$ is a $\tau \times d$ matrix obtained by stacking the data $(x_1, x_2, \allowbreak  \dots, x_\tau)$ and $y_{0, \tau}$ is the $\tau \times 1$ column $(y_1, y_2, \dots, y_\tau)$.

\paragraph{Rolling Regression} Consider the full time horizon for hedging. It is reasonable to assume $\beta$ evolves through time, and the most recent data is more relevant to determine it. Then a naive action is to fix a lookback window $h \ge \tau$ and perform rolling OLS for each $t \in \cT_\tau^n$, and the $\beta$ function is thus given by
\begin{align} \label{eq:rols}
f^*(t, D_{0, t}) = f^*(D_{t-h, t}) = \left(X_{t-h, t}^T X_{t-h, t}\right)^{-1}X_{t-h, t}^T y_{t-h, t}.
\end{align}

\begin{figure*}[h]
\begin{minipage}{\linewidth}    
    \centering
    \begin{subfigure}{0.33\linewidth}
    \centerline{\includegraphics[width=\linewidth]{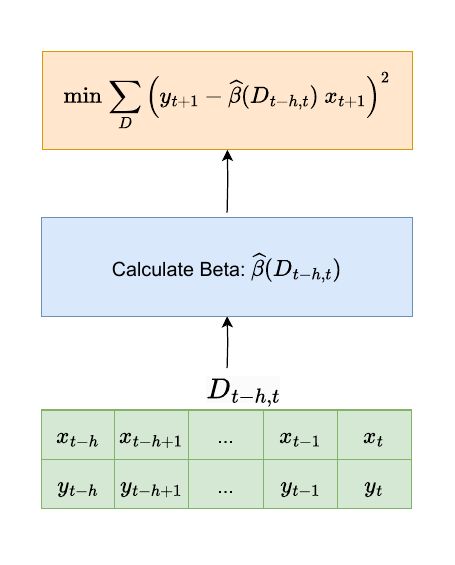}}
    \caption{{General}}\label{fig:general}
    \end{subfigure}\hfill
    \begin{subfigure}{0.33\linewidth}
    \centerline{\includegraphics[width=\linewidth]{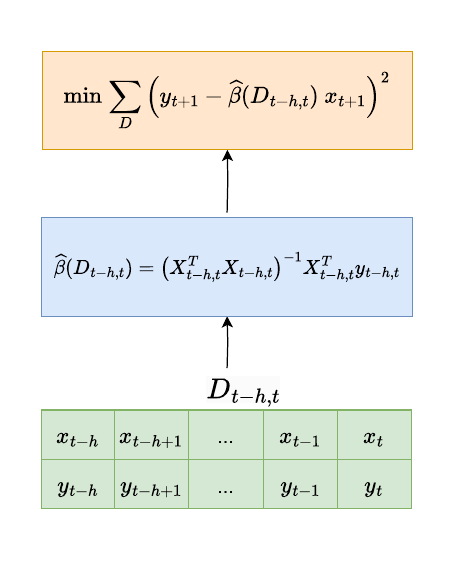}}
    \caption{OLS}\label{fig:ols}
    \end{subfigure}\hfill
    \begin{subfigure}{0.33\linewidth}
    \centerline{\includegraphics[width=\linewidth]{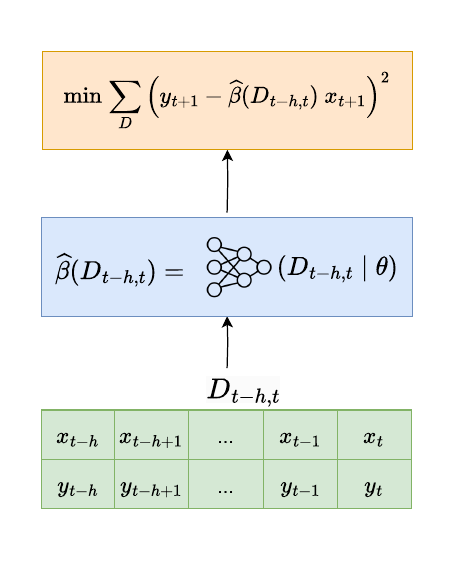}}
    \caption{{NeuralBeta}}\label{fig:neural_beta}
    \end{subfigure}
\end{minipage}
\caption{$\beta$ Estimation Framework}
\label{fig:framework}
\end{figure*}

Figure \ref{fig:ols} shows the process of using rolling OLS to estimate $\beta$. OLS calculates the optimal in-sample $\beta$ and directly uses it as an estimate for the next period, assuming $\beta$ does not change over time.

Two shortcomings of the naive rolling OLS are 1) the selection of lookback window $h$ is somewhat arbitrary, and 2) it ignores data older than $t-h$, while assigning the same importance to each data point within the lookback horizon $(t-h, t]$. So, it might misspecify both long-term and short-term data relevance. One extra step to improve the rolling OLS is replacing OLS with weighted least squares (WLS) by adding a dynamic weighting scheme $w_t = \{w^t_1, \dots w^t_t\} \in \R^t$ to adjust the data importance. Let $W_t = \text{diag}(w_t)$ be the diagonal weight matrix. The estimated $\beta$ then becomes
\begin{align} \label{eq:rwls}
f^*(t, D_{0, t}) = \left(X_{0, \tau}^T W_t X_{0, \tau}\right)^{-1}X_{0, \tau}^T W_t y_{0, \tau},
\end{align}
which aligns with the formula of WLS. Note that rolling OLS \eqref{eq:rols} can be seen as the special case of rolling WLS \eqref{eq:rwls} where $w^t_i = 0$ when $i \le t-h$ and $w^t_i = 1/h$ when $i > t-h$. There are several popular weighting schemes to choose from, such as exponential weights and power laws. In these cases, we have additional parameters to be tuned related to the weighting. Therefore, despite \eqref{eq:rwls} providing flexibility to adjust the data importance, configuring proper weighting schemes $w_t$ for all $t \in \cT_\tau^n$ remains challenging.

\subsection{\textit{NeuralBeta} Model Architecture}

Figure \ref{fig:general} is an illustration on how to perform beta estimation in general. The input is a lookback window of $x$ and $y$. The data is then fed into the model $f$ to estimate beta. The model can have some parameters, like a neural network, or it can be parameter-free, like OLS (Figure \ref{fig:ols}). In this section, we will explain how \textit{NeuralBeta} works, which is represented by Figure \ref{fig:neural_beta}.
\subsubsection{General Framework}
Provided the above discussion, we propose a framework to represent the $\beta$ estimator with a neural network. Let $\theta$ be the parameter set, the \textit{NeuralBeta} model is formulated as
\begin{align} \label{eq:nb_f}
(t, D_{0, t}) \mapsto f(t, D_{0, t}; \theta).
\end{align}

As is shown in Figure \ref{fig:neural_beta}, instead of using some pre-determined formula, \textit{NeuralBeta} model dynamically estimates the target period's $\beta$. More importantly, unlike the regression methods, \textit{NeuralBeta} incorporates numerous parameters that can be properly learned during the training process. In practice, a single \textit{NeuralBeta} mode is utilized across all $x$ and $y$ pairs within an application, so the model is comprehensively trained on every pair, rather than one individual pair. This holistic approach allows \textit{NeuralBeta} model to learn from common patterns that emerge across various pairs.

Given the above discussion on the single target asset case, it is natural to further extend the methodology to handle joint hedging or prediction tasks with multiple time series. A typical example is the factor model. Suppose there are $m \in \N$ time series of target assets $(y_t^i)_{t \in \cT, i \in [m]}$, where $[m]$ simply denotes $\{1,2,\cdots, m\}$, that can be explained by $k \in \N$ time series of common factors and $d-k$ variables idiosyncratic to each asset. For each asset $i \in [m]$, we have similar notations as before for the explanatory variable $x^i_t \in \R^d$, the dataset $D^i=\cup_{s<t} D_{s,t}^i$, and the coefficients of interest with its functional form $\hat{\beta}_{t+1}^i = f^i(t, D_{0,t}^i)$. The objective function \eqref{eq:obj} is extended to joint multiple time series case as

\begin{align} \label{eq:obj_multi}
\min_{(f^i)_{i \in [m]}} \frac{1}{nm} \sum_{t \in \cT_\tau^n, i \in [m]} L\left(y_{t+1}^i, \< \hat{\beta}_{t+1}^i, x_{t+1}^i\>\right).
\end{align}
Note that the argument in objective function \eqref{eq:obj_multi} is a vector of functions $(f^i)_{i \in [m]}$ corresponding to each asset, rather than the single function in \eqref{eq:obj}, but we can still use a single large neural network to represent the vector, i.e., $f^i=f^{i+1}=...=f^m$. The loss function we used in training our models is mean squared error:
$$L\left(y_{t+1}^i, \< \hat{\beta}_{t+1}^i, x_{t+1}^i\>\right) = \left(y_{t+1}^i - \< \hat{\beta}_{t+1}^i, x_{t+1}^i\>\right)^2.$$

We highlight that the \textit{NeuralBeta} model \eqref{eq:nb_f} is a unified framework that not only covers a large range of functional forms of $\beta$, including \eqref{eq:ols}, \eqref{eq:rols}, and \eqref{eq:rwls} discussed above, but also can be easily extended to handle multiple assets and features in higher dimension.

In Section \ref{sec:experiment}, we apply the \textit{NeuralBeta} model framework \eqref{eq:nb_f} to a set of beta estimation tasks that include a single asset scenario \eqref{eq:obj} and multiple assets scenario \eqref{eq:obj_multi}.

\subsubsection{Interpretable Neural Network Architecture}\label{sec:interp}
One concern with using neural networks in beta estimation is the lack of interpretability. To address this issue, we introduce a new output layer inspired by regularized weighted linear regression:
\begin{equation} \label{eq:interp_layer}
     \left(\Sigma^{-1} + X_{t-h, t}^T W_{t-h,t} X_{t-h, t}\right)^{-1}(\Sigma^{-1}  \bm\mu + X_{t-h, t}^T W_{t-h,t} y_{t-h, t})
\end{equation}
 where the weights $W_{t-h,t}$ are positive and $\Sigma$ is a positive-definite matrix. $\bm\mu$ and $\Sigma$ can be interpreted as the mean and covariance of a Gaussian prior over $\beta$. Note that the weights are positive, but unbounded; this is to allow the model to choose how much of the prior to rely on, i.e., small weights lead to more regularization, whereas large weights lead to less regularization.

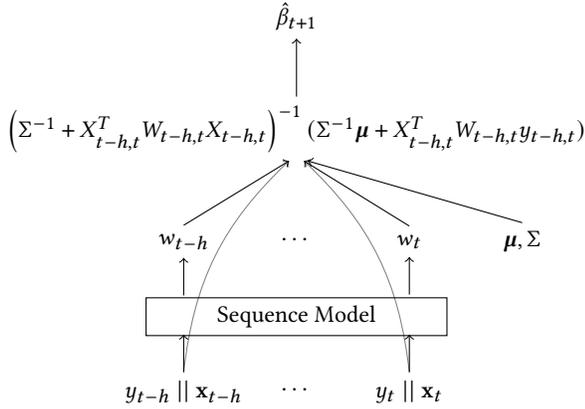
\begin{figure}
\begin{center}
\begin{tikzpicture}[
  every neuron/.style={
    circle,
    minimum size=0.3cm,
    thick
  },
  every data/.style={
    rectangle,
    minimum size=0.4cm,
    thick
  },
]

  \node [align=center,data 1/.try, minimum width=1.5cm] (input-x)  at ($ (2.5, 0) $) {$\bm\mu, \Sigma$};

  \node [align=center,data 1/.try, minimum width=1.5cm] (input-ft)  at ($(input-x) + (-1.5,-2.0)$) {$y_{t}\ ||\ {\mathbf{x}_{t}}$};
  \node [align=center,data 1/.try, minimum width=1.5cm] (input-fdots)  at ($(input-ft) + (-1.5,0)$) {$\cdots$};
  \node [align=center,data 1/.try, minimum width=1.5cm] (input-f1)  at ($(input-fdots) + (-1.5,0)$) {$y_{t-h}\ ||\ {\mathbf{x}_{t-h}}$};

  \node [align=center,data 1/.try, minimum width=4.0cm, draw] (lstm)  at ($(input-fdots) + (-0.0,1.)$) {Sequence Model};

  \node [align=center,data 1/.try, minimum width=1.5cm] (w-t)  at ($(input-ft) + (0, 2.0)$) {$w_{t}$};
  \node [align=center,data 1/.try, minimum width=1.5cm] (input-fdots)  at ($(w-t) + (-1.5,0)$) {$\cdots$};
  \node [align=center,data 1/.try, minimum width=1.5cm] (w-1)  at ($(input-f1) + (0, 2.0)$) {$w_{t-h}$};

    \draw [black,solid,->] ($(input-f1.north)$) -- ($(input-f1.north) + (0,0.5)$);
    \draw [black,solid,->] ($(input-ft.north)$) -- ($(input-ft.north) + (0,0.5)$);

    \draw [black,solid,->] ($(w-1.south) - (0,0.5)$) -- ($(w-1.south) $);
    \draw [black,solid,->] ($(w-t.south) - (0,0.5)$) -- ($(w-t.south)$);

\node [align=center,data 1/.try, minimum width=1.5cm] (rwls)  at ($(lstm) + (0, 2.5)$) {$\left(\Sigma^{-1}  + X_{t-h, t}^T W_{t-h,t} X_{t-h, t}\right)^{-1}(\Sigma^{-1}  \bm\mu + X_{t-h, t}^T W_{t-h,t} y_{t-h, t})$};

\node [align=center,data 1/.try, minimum width=1.5cm] (beta)  at ($(rwls) + (0, 1.5)$) {$\hat{\beta}_{t+1}$};

    \draw [black,solid,->] ($(w-1.north)$) -- ($(rwls.south)-(0.1,0) $);
    \draw [black,solid,->] ($(w-t.north)$) -- ($(rwls.south) + (0.1,0) $);

    \path [black,solid,->, opacity=0.5] ($(input-ft.north)$) edge [bend right=20]  ($(rwls.south) + (0.1, 0)$);
    \path [black,solid,->, opacity=0.5] ($(input-f1.north)$) edge [bend left=20]  ($(rwls.south) - (0.1, 0)$);
    
    \draw [black,solid,->] ($(input-x.north)$) -- ($(rwls.south)+(0.15,0) $);

    \draw [black,solid,->] ($(rwls.north)$) -- ($(beta.south) $);

\end{tikzpicture}
\end{center}
\caption{A diagrammatic representation of \textit{NeuralBeta-Interpretable}. $l$ refers to the lookback window size and $||$ denotes concatenation. Details are in Section \ref{sec:interp}.}\label{fig:stock_embedder}
\end{figure}

In \textit{NeuralBeta}, $\bm\mu$ and $\Sigma$ are global parameters to be trained, and we restrict $\Sigma$ to be diagonal. The weights in \eqref{eq:interp_layer} are the outputs of a sequence model such as a GRU or a transformer. In Figure \ref{fig:stock_embedder}, we show a high-level diagram of our approach which we refer to as \textit{NeuralBeta-Interpretable (NBI)}. Similar to how the neural network in \textit{NeuralBeta} is a replacement of the beta estimator, one interpretation of the "neural" in NBI is that it is implicitly tuning the weights given the context window. We note that this layer reduces the overall expressivity of the neural network; in Section \ref{sec:experiment}, we show that this loss in expressivity has a minor impact on the generalization and can even improve generalization.

Naturally, there remains a certain degree of non-interpretability further upstream, such as understanding the factors that lead certain observations to carry more weights than others. However, this approach takes a meaningful step towards interpretability and generically aid in viewing neural networks as estimators; we leave it to future work to study if the interpretability can be further enhanced.

\section{Experiments}\label{sec:experiment}

In this section, we conduct experiments to estimate $\beta$ in \eqref{eq:beta_estimation} using both synthetic data and market data. We also compare the performance of \textit{NeuralBeta} with rolling OLS and rolling WLS (with exponential weighting scheme) to gain insights into the conditions and mechanisms through which \textit{NeuralBeta} achieves superior performance. To ensure a fair comparison, we tuned the half-life’s of the exponential weights for WLS to achieve
optimal performance in the validation set.

In all the experiments, we use root mean squared error (RMSE) on the predicted $y$ variable as the evaluation metric:
\begin{align} \label{eq:rmse_def}
RMSE(\hat{y}) = \sqrt{\frac{1}{N}\sum_{i=0}^{N}(y_{i}-\hat{y}_{i})^2} \qquad \hat{y}_{i} \coloneqq \< \hat{\beta}_i, x_i\>
\end{align}
The loss function used in training for all the models is mean squared error on the predicted $y$ variable $MSE(\hat{y})$. 
We report the percentage improvement of a model over OLS.

For the sequence model in \textit{NeuralBeta}, we tried GRU \cite{Cho2014GRU} and Attention \cite{vaswani2017attention}. For hyperparameter tuning, we specifically tuned the hidden size (32, 64, 128 and 256), dropout rate (0, 0.25, 0.5), and the lookback window length (64, 128 and 256). If transformer is used as the sequence model, we also tried to tune the patch size (1, 2, 4, and 8) according to \cite{Nie2023PatchTST}. However, this adjustment didn't significantly impact performance, so it is not discussed further in this paper. We trained the models for 100,000 gradient updates using the Adam optimizer \cite{Adam2015} with a learning rate of 1e-4; for evaluation, we use the model with the lowest validation loss, evaluated every 1,000 updates. The hyperparameters are tuned with respect to the validation set, and all the statistics reported in the paper are on the test set. Experimental results are reported in Table \ref{tab:results}; the metric we report is the \% improvement in \textit{RMSE($\hat{y}$)} compared with OLS. 

We implemented \textit{NeuralBeta} in Pytorch \cite{pytorch} and used Pytorch Lightning for training \cite{Falcon_PyTorch_Lightning_2019}.

\subsection{Synthetic Data}

By experimenting on synthetic data, we are able to evaluate the model's performance in different scenarios. More importantly, we know the ground truth $\beta$ in synthetic data, so we can evaluate the beta estimations directly against the ground truth $\beta$, instead of indirectly evaluating on the estimated $y$ variable. It also helps verifying our choice of loss function in market data experiments, i.e., whether it makes sense to evaluate on estimated $y$ when we do not know the ground truth $\beta$.

For synthetic data, we deliberately choose three kinds of representative ground truth $\beta$ coefficients: constant (Section \ref{sec:constant}), stepwise (Section \ref{sec:step}), and cyclical (Section \ref{sec:cyclical}). For each of the synthetic scenarios, we firstly generate the time series of ground-truth $\beta$, then generate the time series $x$ and $y$ accordingly.
\begin{align} \label{eq:syn_step}
x_t & \sim {t}_{10}(0, 1) \\
\epsilon_t & \sim \mathcal{N}(0,1) \\
y_t & = \beta_t x_t + \epsilon_t 
\end{align}
where $t_{\nu}$ refers to a Student's T distribution with $\nu$ degrees of freedom. Each time series is of length 65, and we generated 100,000 samples for each experiment. We use 70\% of the dataset for training, 20\% for validation and 10\% for testing.

\subsubsection{Constant $\beta$}\label{sec:constant}
Constant $\beta$ corresponds to the simplest case, where the response variable has a time-invariant relation with the explanatory variable:
\begin{align} \label{eq:syn_const}
\beta_t \equiv c, \quad c \in \R.
\end{align}
We sample $c$ from $\mathcal{N}(1,1)$ as the constant $\beta$ throughout the whole period for each sample. 
Theoretically, the optimal solution here is the posterior mean (derived via Bayesian linear regression):
\begin{align} \label{eq:bayesian}
\left(\Lambda_0 + X_{t-h, t}^T X_{t-h, t}\right)^{-1}(\Lambda_0  \bm\mu_0 + X_{t-h, t}^T y_{t-h, t})
\end{align}
where $\Lambda_0$ and $\bm\mu_0$ denote the prior mean and precision matrix of $\beta$. In this case, the correct $\Lambda_0$ and $\bm\mu_0$ to use are all $1$ since $\beta$ is generated from $\mathcal{N}(1,1)$. Nonetheless, it is non-trivial to test out whether the \textit{NeuralBeta} model can converge to the optimal solution \eqref{eq:bayesian}.

\subsubsection{Stepwise $\beta$}\label{sec:step}
This is the simplest case for time-varying beta. It also corresponds to market regime shifts where the $\beta$ coefficient between two assets stays constant for a certain period then jumps to a new level. This scenario can test the adaptivity of \textit{NeuralBeta} model to sudden changes in the market. We have 
\begin{align} 
\beta_t = \sum_{i=1}^{n} \beta_i \one_{\cT_i}(t),
\end{align}
for $n$ constants $\beta_i \in \R$ and disjoint time intervals $\cT_i$ that partition the entire horizon $\cT$.  For our experiments, we chose $n=2$ and randomly picked a location in the lookback window as the jump position. The $\beta$ coefficients before and after the jump are also generated from $\mathcal{N}(1,1)$.

\subsubsection{Cyclical $\beta$}\label{sec:cyclical}
Certain financial time series demonstrate seasonality or other cyclical behaviors due to various reasons, such as agricultural production, business cycles, retail sales, etc. Classic models for cyclical patterns like decomposition models \cite{cleveland1990stl} and seasonal autoregressive moving average model \cite{box2015time} often involve steps to separate trend and seasonality factors before estimation, which requires attentive treatment. Unlike existing models, the \textit{NeuralBeta} model is able to dynamically capture the cyclical pattern of the $\beta$ coefficient without further modification to the model. We test \textit{NeuralBeta} on a simple example of sinusoidal $\beta$ in the form of
\begin{align} \label{eq:syn_cyclic}
\beta_t & = \sin(\beta_0 + c t) \\
\beta_0 & \sim N(0,1) \\
c & \sim U(4,32)
\end{align}
where $U(a,b)$ refers to a uniform distribution between $a$ and $b$.
The parameters are chosen such that each sample contains at least one half period of the sinusoidal wave.

\subsubsection{Performance Analysis}
We test \textit{NeuralBeta} on the above synthetic datasets against the benchmarks (rolling OLS and rolling WLS). In Table \ref{tab:results}, we can see in all three cases of synthetic $\beta$, \textit{NeuralBeta} significantly outperforms OLS and WLS. Even in the constant $\beta$ case where OLS is very close to the theoretically optimal solution, \textit{NeuralBeta} still outperforms. More importantly, the performance of interpretable \textit{NeuralBeta} and non-interpretable ones are almost identical, and interpretable \textit{NeuralBeta} with attention mechanism outperforms except for the cyclical $\beta$ case. This shows that the interpretable architecture does not sacrifice performance while providing more transparency.

\begin{table}[!b] 
\small
\begin{tabular}{lcccccc}
\toprule
     & \multicolumn{2}{c}{Baseline}
     & \multicolumn{2}{c}{GRU}
     & \multicolumn{2}{c}{Attn}
\\
    \cmidrule(lr){2-3}
    \cmidrule(lr){4-5}
    \cmidrule(lr){6-7}
         & OLS & WLS & NB & NBI& NB& NBI   \\ 
\midrule
Constant & $0.00$ & $-0.05$   & $-0.04$ & $-0.01$  & $-0.13$  & $\bm{0.01}$ \\
Stepwise & $0.00$ & $20.21$ & $21.80$ & $21.98$  & $21.65$  & $\bm{22.22}$ \\
Cyclical   & $0.00$ & $17.84$ & $\bm{22.54}$ & $21.95$  & $21.10$  & $20.84$ \\
\midrule
Univariate & $0.00$ &  $0.12$  & ${0.22}$ & $0.26$  & $0.31$  & $\bm{0.40}$ \\      
Multivariate & $0.00$ & $0.34$ & $-2.02$ & $0.84$ & $-4.49$  & $\bm{0.95}$ \\ \bottomrule
\end{tabular}
\caption{\% Improvement against benchmark (OLS). The best result for each scenario is in bold.}
\label{tab:results}
\end{table}

\begin{figure}[!b]
 \centerline{\includegraphics[width=0.9\linewidth]{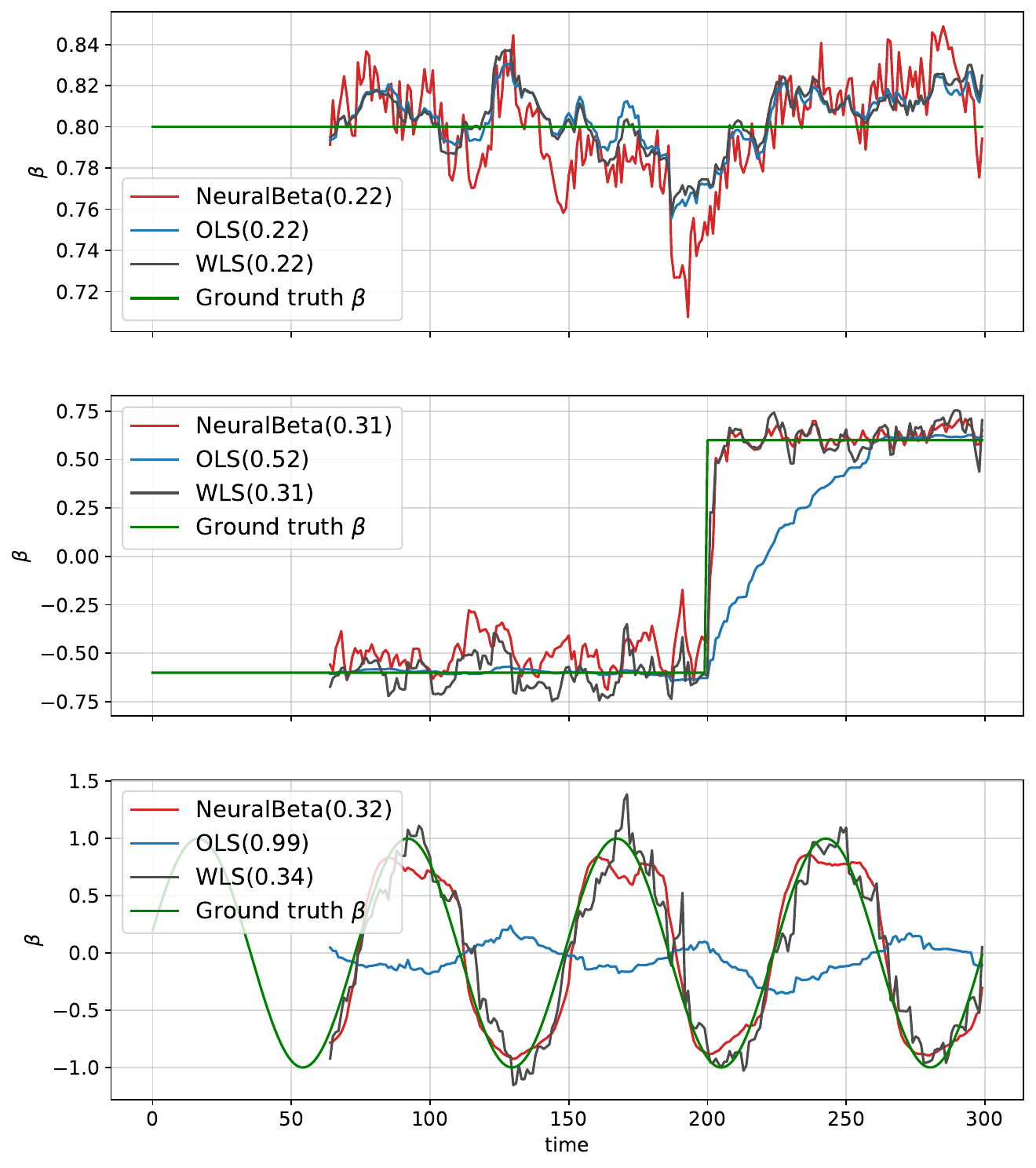}}
    \caption{Estimations of $\beta$. \textit{RMSE($\hat{y}$)} of each model is shown in parentheses in the legend.}
    \label{fig:beta_estimations}
\end{figure}

Figure \ref{fig:beta_estimations} presents examples of the ground truth $\beta$ values alongside the model estimations for the three different scenarios: constant $\beta$, stepwise $\beta$ and cyclical $\beta$, from top to bottom respectively. Three models - OLS, WLS with exponential weights, and \textit{NeuralBeta} - were applied to these datasets. Note that, to ensure a fair comparison, we tuned the half-life's of the exponential weights for WLS to achieve optimal performance, and selected the best-performing \textit{NeuralBeta} model for each dataset. In the constant $\beta$ scenario, equation \eqref{eq:bayesian} is the optimal solution in theory. \textit{NeuralBeta}'s performance is slightly better than OLS in terms of \textit{RMSE($\hat{y}$)}, which proves that \textit{NeuralBeta} is able to identify the best strategy in this simple scenario. For the stepwise $\beta$ scenario, both \textit{NeuralBeta} and WLS significantly outperform OLS and perform almost identically to each other. Here the WLS model uses a half-life of 2, which means that it heavily weighs the two most recent data points - an effective approach for stepwise $\beta$. \textit{NeuralBeta} automatically identifies and applies this strategy. In the cyclical $\beta$ scenario, \textit{NeuralBeta} outperforms both OLS and WLS. While WLS, with a half-life of 1, can track the sine wave's trend, \textit{NeuralBeta} provides smoother and more accurate estimations, while maintaining the same level of responsiveness. This demonstrates \textit{NeuralBeta}'s superior capability in handling dynamic and complex $\beta$ patterns.

\begin{figure}[!bt]
    \centering
    \includegraphics[width=\linewidth]{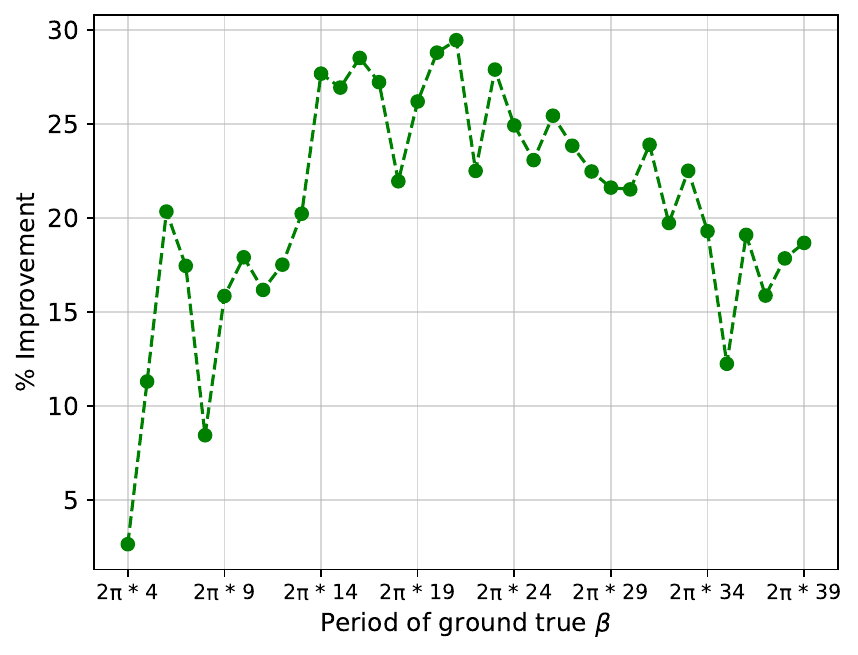}
    \caption{Improvements compared with OLS for cyclical $\beta$ across different periods. \textit{NeuralBeta} achieves best performance when $\beta$ changes at a moderate rate.}
    \label{fig:sine_improvement}
\end{figure}

Figure \ref{fig:sine_improvement} shows the improvement of \textit{NeuralBeta} model compared with OLS on cyclical $\beta$ with different periods of the sine wave. The x-axis denotes the period of the ground truth sinusoidal wave used as $\beta$ in our synthetic data, while the y-axis represents the improvement of \textit{NeuralBeta} over OLS in \% terms. \textit{NeuralBeta} achieves the most significant improvement over OLS when the period of the sinusoidal $\beta$ is moderate. When the period is either very low (indicating fast changes in $\beta$), or very high (indicating slow changes in $\beta$), \textit{NeuralBeta}'s advantage over OLS diminishes, though it consistently outperforms OLS across all periods. This observation indicates that \textit{NeuralBeta} model excels when $\beta$ changes at a moderate rate.

\subsubsection{Comparing \textit{RMSE($\hat{y}$)} and  \textit{RMSE($\hat{\beta}$)}}
\begin{figure}[!bt]
    \centering
    \begin{subfigure}{0.8\linewidth}
    \centerline{\includegraphics[width=\linewidth]{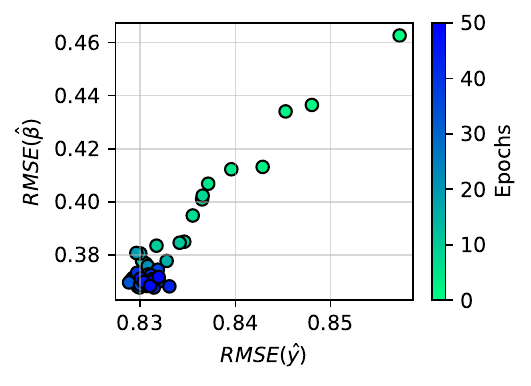}}
    \caption{{Attention}}\label{fig:attention}
    \end{subfigure}\hfill
    \begin{subfigure}{0.8\linewidth}
    \centerline{\includegraphics[width=\linewidth]{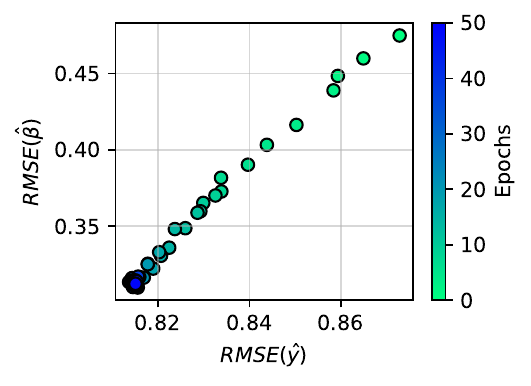}}
    \caption{{GRU}}\label{fig:gru}
    \end{subfigure}
    \caption{\textit{RMSE($\hat{y}$)} and \textit{RMSE($\hat{\beta}$)} are highly correlated.}
    \label{fig:correlated_errors}
\end{figure}

One important reason why we want to do synthetic experiments instead of testing the model directly on market data is that we want to examine the correlation between \textit{RMSE($\hat{y}$)} and \textit{RMSE($\hat{\beta}$)}. Since in real data we do not know the ground truth $\beta$, it is impossible to compute \textit{RMSE($\hat{\beta}$)} directly. If we can verify that errors on $\hat{y}$ and on $\hat{\beta}$ are highly correlated, we can confidently use the error on the predicted $y$ as a reliable proxy for the model's performance. Figure \ref{fig:correlated_errors} shows \textit{RMSE($\hat{y}$)} and \textit{RMSE($\hat{\beta}$)} across various epochs of the same model in the cyclical $\beta$ scenario. We use attention as the underlying sequence model in Figure \ref{fig:attention} and GRU in Figure \ref{fig:gru}. The analysis is limited to the first 50 epochs, as performance stabilizes beyond this point. In both figures, \textit{RMSE($\hat{y}$)} and \textit{RMSE($\hat{\beta}$)} appear to be highly correlated, which further justifies the validity of our approach. This correlation supports the use of \textit{RMSE($\hat{y}$)} as an effective measure for evaluating the quality of $\beta$ estimators when the ground truth $\beta$ is unavailable.

\subsubsection{Analyzing Weights from NBI}
\begin{figure}[!bt]
    \centering
    \includegraphics[width=\linewidth]{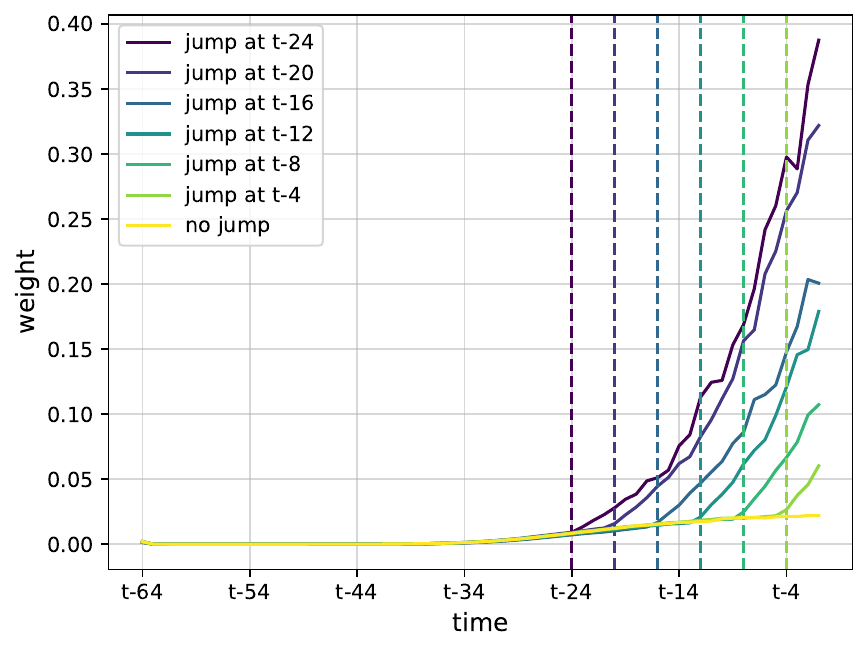}
    \caption{Weights assigned by NBI for stepwise $\beta$ when $\beta$ jumps at different positions.}
    \label{fig:step_weights}
\end{figure}

In Figure \ref{fig:step_weights}, we present the weights generated by \textit{NeuralBeta-Interpretable} (NBI) in the stepwise $\beta$ scenario across a lookback window of length 64. We selected different positions where the ground truth $\beta$ jumps from 2 to 0, generated 1,000 samples for each position, and calculated the average weights assigned by NBI. The results show a significant increase in weights for data points following a jump, while the weights before the jump remain close to 0. This demonstrates \textit{NeuralBeta}'s ability to swiftly detect and adapt to changes in $\beta$ .

With the above experiments, we show \textit{NeuralBeta}'s ability to dynamically adapt to regime shifts, and to precisely capture cyclical market moves. In the next section we will test  
how the \textit{NeuralBeta} model can possibly handle more complicated market data.

\subsection{Market Data}
The market dataset for our experiments covers the time horizon from 2010-01-01 to 2023-12-31. We use the daily return series of the S\&P 500 index, the size factor index, the value factor index, and all S\&P 500 components. We set 2010-01-01 to 2017-12-31 as the training period, 2018-01-01 to 2019-12-31 as the validation period, and 2020-01-01 to 2023-12-31 as the test period. As member companies in S\&P 500 varies over time, research on investment strategies often accounts for these changes to avoid look-ahead bias during backtesting. However, since our primary focus is on the predictive power of $\beta$ coefficient estimators instead of backtesting strategies, we do not need to track the change of index components. For simplicity, we use a snapshot S\&P 500 components as of 2024-05-01, and keep only those stocks with price histories available from 2010/01/01. This results in a dataset of 468 stocks used for our analysis.

Specifically, our experiments cover both the univariate and multivariate scenarios. For the univariate scenario, we aim to calculate the CAPM $\beta$ for each stock in S\&P 500. For the multivariate case, we aim to calculate the factor $\beta$'s for the same universe with respect to the market factor, the size factor (R2FSF Index), and the value factor (RAV Index), which is similar to the setup in the Fama-French three-factor model.

\subsubsection{S\&P 500 Components with CAPM}\label{sec:uni_market}

In this experiment, we use the daily return series of the S\&P 500 index (SPX Index) along with the daily return series of its individual components. The "Univariate" entry in Table \ref{tab:results} shows the performance of different models on this dataset.

From Table \ref{tab:results}, we can see that using attention as the underlying sequence model is slightly better than using GRU, and the interpretable architecture is slightly better than general architecture in terms of test performance. This again indicates that incorporating an interpretable architecture, although introduces a restriction to the capacity, does not compromise the model's performance compared to the general, non-interpretable version. 

\begin{figure}[!bt]
    \centering
    \includegraphics[width=0.9 \linewidth]{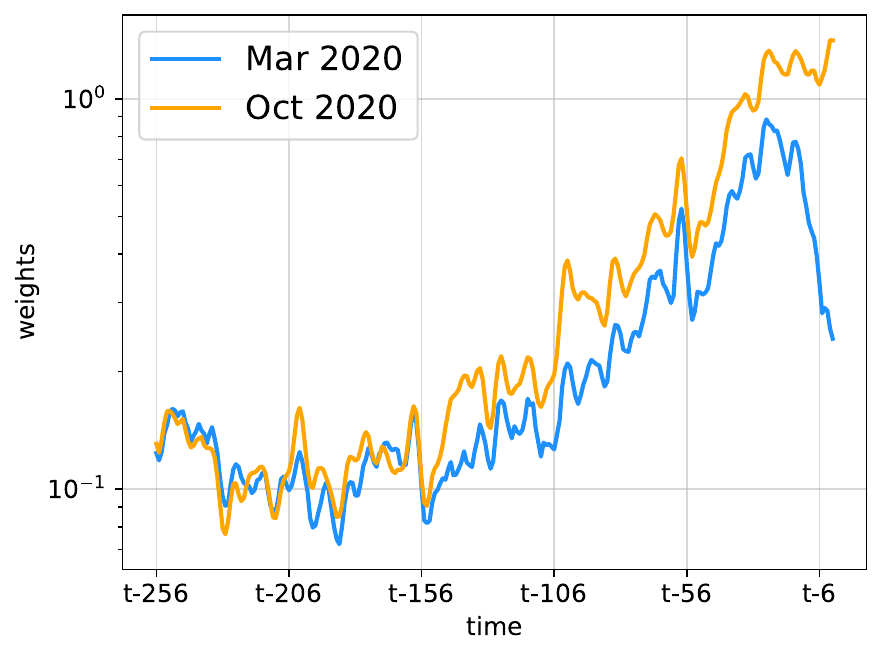}
    \caption{Average weights in log scale in the lookback window. March 2020 puts much less weights on recent data compared with October 2020.}
    \label{fig:weights}
\end{figure}

To further investigate the model's behavior during different market conditions, we plotted the average weights assigned to each lag in the lookback window of 256 days for two distinct periods: the beginning of COVID-19 (March 2020) and October 2020. As shown in Figure \ref{fig:weights}, during the onset of the pandemic in March 2020, the model assigned significantly lower weights to the most recent data points. This likely reflects the extreme volatility present at that time which makes the data at that time less reliable for prediction purposes. By reducing the weights on these recent data points, the model aimed to mitigate the impact of short-term anomalies on beta estimation. Conversely, in October 2020, as market conditions began to stabilize, the model assigned higher weights to the most recent data points, which indicates a restored confidence in the relevance of recent data. It is also noteworthy that the log weights in October 2020 show a linear trend in the latter half of the lookback window, indicating exponential growth in the weights. This is significant, as the exponential weighting scheme, which is recognized by researchers as an effective solution, is identifies by \textit{NeuralBeta} without human intervention.

\begin{figure}[!bt]
    \centering
    \includegraphics[width=\linewidth]{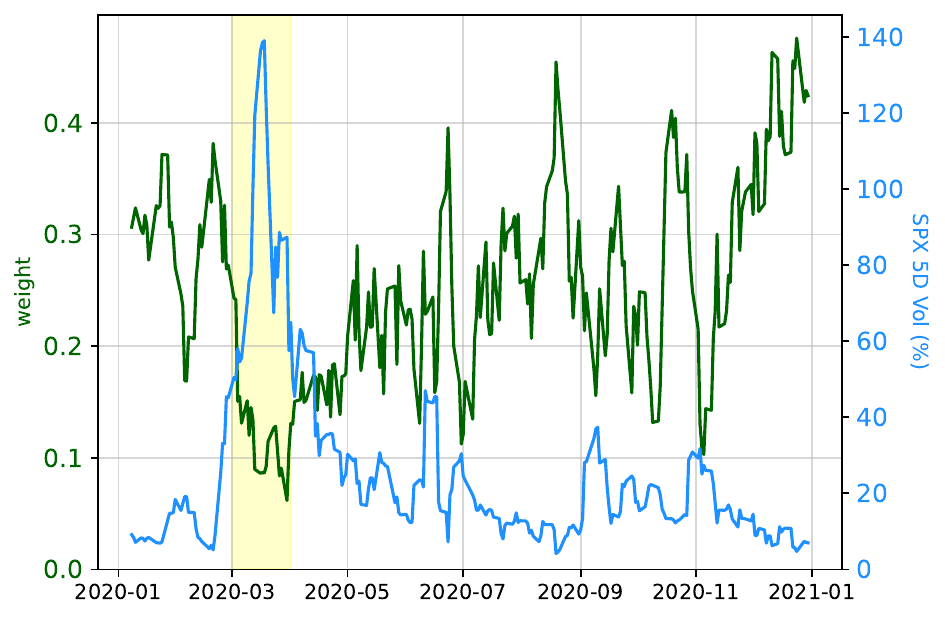}
    \caption{Average weights in 2020 with SPX 5-day volatility. Weights are lower during periods of high volatility.}
    \label{fig:weights_by_date}
\end{figure}

Additionally, Figure \ref{fig:weights_by_date} shows the average weights per day in the S\&P500 universe, along with the 5-day volatility of SPX. Generally, the model tends to assign lower weights during periods of high volatility and higher weights when volatility subsides. By examining the weights, we can see the model's ability to adapt to changing market conditions and produce robust estimates.

\subsubsection{S\&P 500 Components with Factors}

As an extension of our experiments, we utilized the daily return series of the SPX index, along with size and value indices, to perform multivariate beta estimation on S\&P 500 components. This experiment allows us to assess the model's ability to handle multiple factors instead of just one factor. The performance of the multivariate model is shown in the "Multivariate" entry in Table \ref{tab:results}. In this case, the interpretable architecture with Attention performs the best, while the non-interpretable \textit{NeuralBeta} models performs slightly worse than the benchmark. This experiment shows \textit{NeuralBeta}'s potential in multi-factor models.

\subsection{Hyperparameter Tuning}

\begin{figure*}[tb]
 \centerline{\includegraphics[width=\linewidth]{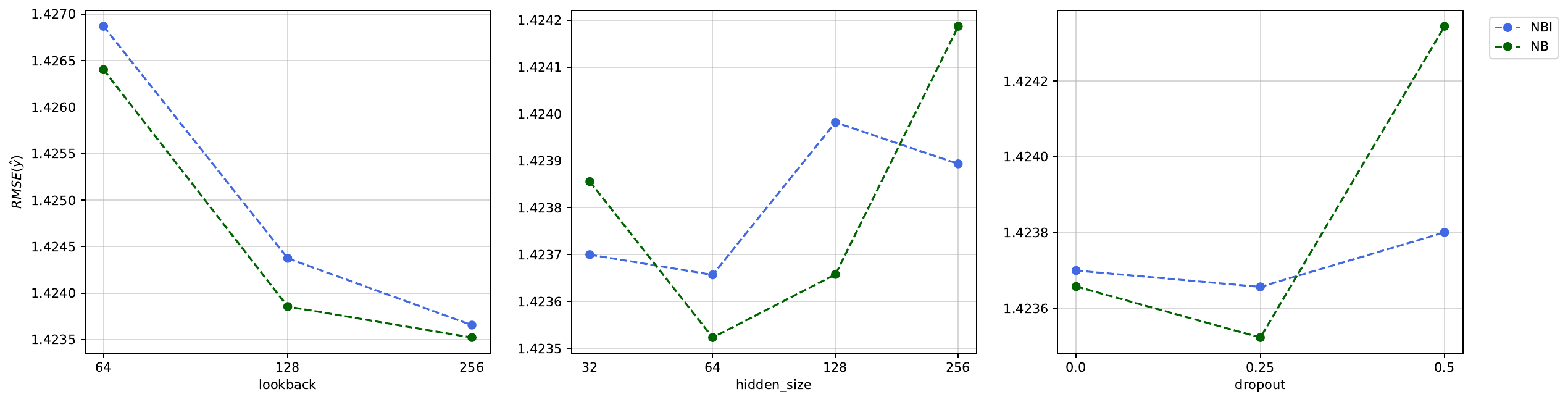}}
    \caption{\textit{RMSE($\hat{y}$)} of NB and NBI for different hyper-parameter values. NBI is more robust across different configurations.}
    \label{fig:hypertune_curve}
\end{figure*}

To understand how the performance changes with different hyperparameter settings, in Figure \ref{fig:hypertune_curve}, we plot the validation \textit{RMSE($\hat{y}$)} of \textit{NeuralBeta-Interpretable} (NBI) and \textit{NeuralBeta} (NB) with Attention for the experiment in \ref{sec:uni_market} across a range of values for lookback, hidden size, and dropout. For each parameter value, we select the best-performing model from all possible parameter combinations and present the results in the plots.

The performance analysis shows that a longer lookback window leads to better results due to the increased information it provides. A medium hidden size and dropout rate are preferred. Although the interpretable version (NBI) performs slightly worse than the non-interpretable version on the validation set in this experiment (NBI performs better on the test set, as is shown in Table \ref{tab:results}), it demonstrates greater robustness, with its performance showing less variability across different parameter values. This robustness is likely because of NBI's architectural design, which inherently incorporates the linear regression formula and reduces the complexity of the function needed to learn from raw data. NB, on the other hand, must learn them from scratch, which could lead to higher variance across different configurations.


\section{Conclusion}\label{sec:conclusion}

In this paper, we presented \textit{NeuralBeta}, a novel approach to beta estimation using deep neural network. This model effectively addresses several challenges associated with traditional beta estimation techniques, particularly the challenge with dynamic $\beta$ values and model transparency. 
Further, we also developed an interpretable neural network, \textit{NeuralBeta-Interpretable}, which we find improves not only transparency but also performance.

We conducted extensive experiments on both synthetic and market data to validate the efficiency and robustness of \textit{NeuralBeta}. The results demonstrated superior performance across diverse scenarios compared with benchmark methods. We also provided examples of the weights produced by the interpretable version of \textit{NeuralBeta}. These examples illustrate how \textit{NeuralBeta} derives its predictions and offer users insight into the model's mechanism by examining the distribution of weights.

Beyond beta estimation, this model can be extended to other financial settings that assume linear relationships, such as calculating an option's delta in options pricing. Given the prevalence of linear assumptions in financial modeling, \textit{NeuralBeta}'s generality, interpretability, and practicality make it a powerful tool for many financial applications.

\bibliographystyle{ACM-Reference-Format}
\bibliography{sample}

\end{document}